\documentclass[12pt,twoside]{article}   
\usepackage[textwidth = 18cm]{geometry}
\usepackage{caption}
\usepackage{amssymb,amsmath}
\usepackage{pdfpages}

\usepackage[super,square,sort,comma]{natbib}
\usepackage{enumitem}

\usepackage{ragged2e}
\usepackage{fancyhdr}		

\usepackage[export]{adjustbox}
\usepackage{mwe}
\usepackage{framed}
\usepackage{mdframed}
\usepackage{caption}
\usepackage[strict]{changepage}
\usepackage{longtable}
\usepackage{array}
\usepackage{booktabs}
\usepackage{graphicx}
\usepackage{framed}
\usepackage{tabularx}
\usepackage{array}
\usepackage{makecell}
\usepackage{caption}
\usepackage{floatrow}

\DeclareFloatFont{tiny}{\tiny}

\usepackage{etoolbox}
\makeatletter
\usepackage{enumitem}
\setlist{nosep}
\usepackage{sectsty}

\usepackage{etoolbox}
\usepackage{tabularray}
\usepackage{booktabs}

\usepackage{framed}
\usepackage{mdframed}
\usepackage[strict]{changepage}

\usepackage{graphicx}
\usepackage{framed}
\usepackage{tabularx}
\usepackage{array}
\usepackage{makecell}
\usepackage{floatrow}
\DeclareFloatFont{tiny}{\tiny}

\usepackage{textcomp}
\usepackage{etoolbox}
\makeatletter

\usepackage{sectsty}

\usepackage{abstract}

\usepackage{tabularray}
\usepackage{booktabs}
\usepackage{caption}

\newcommand{\captionv}[3]{\begin{center}\parbox{#1cm}{\caption[#2]{{\sf #3}}}
        \end{center}}

\usepackage[section]{placeins}   %
\setlist[itemize]{leftmargin=1em}
\usepackage{caption}
\usepackage{xcolor}

\usepackage{enumitem}
\usepackage[T1]{fontenc}
\usepackage{graphicx}
\makeatletter \renewcommand\@biblabel[1]{$^{#1}$} \makeatother
\newcommand{\cen}[1]{\begin{center} #1 \end{center}}

\usepackage{caption} 

\usepackage{lineno}

\usepackage{hyperref}
\hypersetup{ colorlinks,
    citecolor=blue,
    filecolor=blue,
    linkcolor=blue,
    urlcolor=blue
}
\usepackage{titlesec}
\titlespacing*{\subsection}{0.5pt}{1\baselineskip}{0.1\baselineskip}
\mdfdefinestyle{boxcaption}{innerleftmargin=0.45cm,innerrightmargin=0.45cm}

\titlespacing*{\subsubsection}{0.5pt}{1\baselineskip}{0.1\baselineskip}
\titlespacing*{\section}{0.5pt}{1\baselineskip}{0.1\baselineskip}

\usepackage{xcolor}

\definecolor{gray}{rgb}{0.6,0.6,0.6}
\definecolor{red}{rgb}{0.85,0,0}
\definecolor{green}{rgb}{0,0.85,0}
\definecolor{blue}{rgb}{0,0,0.85}
\definecolor{beige}{rgb}{0.92,0.87,0.78}
\usepackage[all]{hypcap}    
\usepackage{setspace}

\begin{document}

\cen{\sf {\LARGE {\bfseries ORCHID: A Semantic Extension of mCODE for Radiomics and Dosiomics in Radiotherapy} }

Odette Rios-Ibacache$^{1}$,
Gaetan Butault$^{2}$,
Amal Zouaq$^{2}$ \&
John Kildea$^{1}$

$^{1}$Medical Physics Unit, McGill University, Montreal, QC, Canada\\
$^{2}$Department of Computer Engineering and Software Engineering, Polytechnique Montreal, Montreal, QC, Canada}
\vspace{-3mm}
\pagenumbering{roman}
\setcounter{page}{1}
\pagestyle{plain}

\noindent \textbf{keywords:}  \textit{mCODE, ontology, data standards, semantic interoperability, radiotherapy, radiomics, dosiomics}\\
Corresponding author. email: odette.riosibacache@mail.mcgill.ca

\vspace{-3mm}
\begin{center}
    \textbf{Abstract}
\end{center}
\vspace{-4mm}
\noindent{\bf Background:} The scattered nature of health data limits Artificial Intelligence (AI) from integrating multimodal data sources, such as Electronic Health Records (EHRs), genomics, and radiomics, to automate patients' outcome assessments. Additionally, institution-specific imaging and data-storage practices further hinder data sharing across healthcare institutions. Although the Minimal Common Oncology Data Elements (mCODE\textsuperscript{\textregistered}) standard provides a structured foundation for oncology data, quantitative radiomics and dosiomics also require explicit representation of imaging acquisition and reconstruction parameters, feature-extraction settings, and dose-derived features in their clinical and treatment contexts.\\ 
{\bf Purpose:} This project aimed to develop ORCHID (Ontology for Radiomics, Health, and Imaging Data), an ontology built on the mCODE framework and designed to support the standardized semantic representation of clinical oncology data alongside medical imaging, radiomics, and dosiomics information, with particular emphasis on data elements relevant to quantitative imaging and radiotherapy research.\\
{\bf Methods:} We first identified essential data elements for medical imaging, radiomics, and dosiomics and used them to propose a possible mCODE extension linking these data to patients’ EHRs. A literature review identified key parameters affecting radiomics feature extraction and reproducibility. Next, we developed our ontology following the five stages of METHONTOLOGY: domain specification, ontology reuse, conceptualization, implementation, and evaluation. Our ontology was built using the Web Ontology Language (OWL) software Protégé, and logical consistency and quality assurance were evaluated using the HermiT reasoner and ROBOT (a command-line tool for ontology development), respectively. Finally, we created ten random instances with clinical information taken from mCODE-based Synthea Synthetic Data, and queried the data using SPARQL to verify successful information retrieval. \\
{\bf Results:} The resulting ontology, ORCHID, extends the mCODE framework with 99 data elements in medical imaging, radiomics, and dosiomics. When combined with imported ontologies, our ontology comprises 152,129 classes, 352 object properties, 47 data properties, and 216,754 logical axioms. Concepts were reused from established biomedical and oncology ontologies, including SNOMED CT, ICD-10, RadLex, O-RAW, the Operational Ontology for Oncology (O3), the Sequence Ontology (SO), and the Units of Measurement Ontology. No logical inconsistencies were found. \\
{\bf Conclusions:} This study delivers an ontology compatible with mCODE and established oncology ontologies, which unifies biomedical standards and multimodal oncology data. It offers a foundation for interoperable and scalable oncology data research, data sharing, and analysis.

\setlength{\baselineskip}{0.6cm}      
\pagenumbering{arabic}
\setcounter{page}{1}
\pagestyle{fancy}

\vspace{-4mm}
\setstretch{1}
\section{Introduction}
Artificial Intelligence (AI) has the potential to transform oncology and medical physics by integrating multisource data—including electronic health records (EHRs), medical images, and radiotherapy (RT) information—to enable automated clinical assessments, image-based tumor targeting, and personalized RT treatment planning \cite{Krishnamurthy2022-zn, Bajwa2021-ds, Rios-Ibacache2026-sv, Mayerhoefer2020-td, Kang2020-gn}. Radiomics and dosiomics, which derive quantitative features from medical images and radiation dose distributions, respectively, provide complementary sources of potential biomarkers for predictive modeling research \cite{Park2018-oe,Dominguez2023-cc,Ching2023-ic}. However, these features are sensitive to variations in image acquisition, scanner characteristics, preprocessing, segmentation, and feature-extraction settings \cite{Stamoulou2022-cm, Zhao2021-hu, Sedlakova2023-iq, Espinasse2020-yw}. These variations, together with differences in patient cohorts, can limit cross-institutional reproducibility, introduce bias, and hinder the generalizability of AI models \cite{Xu2025-qc, LINTONREID2026107165, Soliman2022-rx,Stamoulou2022-aq}. Addressing these challenges requires standardized representations not only of quantitative features, but also of their source data, derivation parameters, and relationships with clinical and treatment information.

Within institutions, EHRs manage real-world data (RWD), including pathology reports, laboratory results, and medication records \cite{Eichler2019-zm}. However, heterogeneous recording practices and lack of interoperability hinder the exchange of EHR data and limit inter-institution research \cite{Tsai2020-dw,Sedlakova2023-om, Abdelouahid}. Data standards are essential for consistent representation and reuse of clinical information \cite{Gliklich2019-vu}. Without interoperable infrastructures, research may remain restricted to single-center cohorts, limiting the generalizability of downstream AI applications, including digital twins \cite{McDonald2001-ya,Hughes2023,Katsoulakis2024-fn}. Moreover, the growing complexity of clinical, imaging, and genomic data further motivates structured, machine-readable approaches to multimodal integration \cite{Linder2021-kt,Mumtaz2023-vy,Fessele2018-dm}.

The Minimal Common Oncology Data Elements (mCODE\textsuperscript{\textregistered}) standard, developed with support from the American Society of Clinical Oncology (ASCO), defines a core set of structured oncology data elements to facilitate consistent representation and exchange of cancer-related information \cite{Osterman2020-om}. However, mCODE is a data standard rather than an ontology, and it does not provide a dedicated representation of the imaging, radiomics, and dosiomics information required for medical physics research. Ontologies complement data standards by formally defining concepts and relationships to support machine-readable representation and retrieval\cite{Guarino2009-ku, article_Advantages}. Recent work, including the Cancer Image Europe Hyperontology, has advanced semantic integration across oncology and medical imaging \cite{El_Ghosh2026-je}. However, representing radiomics extraction settings and dose-derived features alongside relevant imaging acquisition, reconstruction, processing, and RT information remains important for quantitative medical physics research.

In this study, we developed ORCHID: Ontology for Radiomics, Health, and Imaging Data, an mCODE-based ontology with an extension for medical imaging, radiomics, and dosiomics. The extension represents imaging acquisition and reconstruction parameters, radiomics feature-extraction settings, and dose-derived features linked to RT information. This work extends mCODE to represent medical imaging, radiomics, and dosiomics within a shared clinical oncology context, supporting cross-domain information retrieval and providing a foundation for investigating interoperability and reproducibility in RT research.
\vspace{-5mm}
\section{Materials and Methods}

To establish a semantically interoperable framework, our methodology encompasses three sequential stages: first, defining the multimodal mCODE extension via literature review of imaging and feature extraction parameters; second, engineering an mCODE-based OWL ontology; and third, deploying a web-based documentation browser to facilitate user interaction and ontology exploration.
\vspace{-5mm}
\subsection{mCODE Extension Development}
mCODE Standard Trial of Use (STU) 4 provided the minimal core set of structured data elements required for our ontology, including genomics-related components, that inform cancer assessment and treatment decisions across six core groups: \textit{Patient Information, Disease Characterization, Health Assessment, Cancer Treatments, Outcomes}, and \textit{Genomics}\cite{Osterman2020-om, HL7_mCODE_Genomics}. To address the lack of dedicated representations for medical imaging, radiomics, and dosiomics, we proposed an extension comprising three additional groups: Medical Imaging, Radiomics, and Dosiomics. Dosiomics refers to quantitative features derived from RT dose distributions \cite{Liang2019-ov}.

We analyzed the elemental structure of mCODE and designed our extension by linking the \textit{Cancer Patient} in a similar approach to mCODE's \textit{Genomics Report}, given that genomics and radiomics are both considered cancer biomarkers \cite{He2024-jw}. Thus, we stated that each \textit{Cancer Patient} should have a \textit{Radiomics Report} and a \textit{Dosiomics Report} that would describe their tumor and disease. The extension links the three groups to the \textit{Cancer Patient} entity, enabling the representation of diagnostic and RT images, radiomics and dosiomics reports, and their relationships to clinical and treatment information. Because quantitative imaging and dose-derived features depend on data acquisition and processing, the extension must also ideally capture relevant imaging acquisition, reconstruction, preprocessing, and feature-extraction parameters.

\begin{figure}[h]
\vspace{-11mm}
\centering
{\captionv{12}{Short title - can be blank}{\small METHONTOLOGY-based workflow for oncology ontology engineering. The five-step process covers domain scope definition, terminology reuse, conceptual modelling, Protégé implementation, and evaluation via reasoning, quality assurance, and SPARQL queries.\label{fig:methontology}}}
\includegraphics[width=0.85\linewidth]{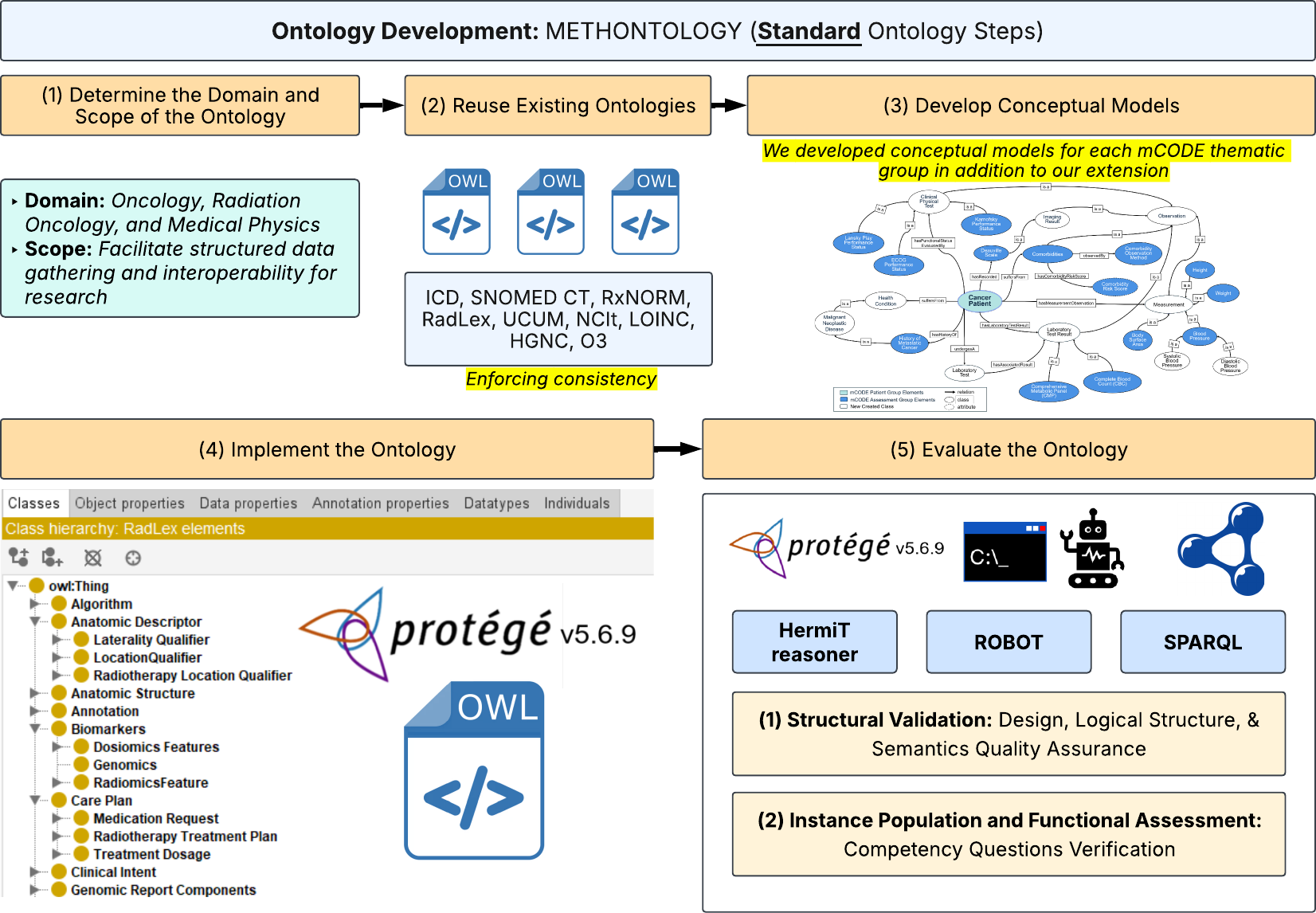}
\end{figure}
\vspace{-5mm}
\subsubsection{Elements Identification}

To identify the minimum data elements that would contain our proposed extension to mCODE, we formulated a guiding question: \textit{What information must be captured to understand how quantitative features were generated, identify methodological factors that may influence their values, and support meaningful comparisons across studies?} We prioritized relevant parameters related to image acquisition, reconstruction, preprocessing, and feature extraction, while considering the feasibility of extracting data elements from DICOM files. This approach aimed to improve feature interpretability, support reproducibility, and help identify potential sources of methodological bias and variability. The Image Biomarker Standardisation Initiative (IBSI) recommendations informed the selection of radiomics parameters and alignment with existing standardization practices \cite{Zwanenburg2020-uu, Hatt287}.

To address our question, we conducted a narrative literature review\cite{Sukhera2022-nr} of studies published between 2016 and 2026 on radiomics and dosiomics reproducibility, robustness, standardization, and harmonization. We specifically included articles that explicitly reported or evaluated imaging and radiomics acquisition parameters as sources of variability. PubMed and Google Scholar were searched using combinations of “radiomics,” “dosiomics,” “standardization,” “harmonization,” “variability,” “feature extraction,” “reproducibility,” “imaging protocols,” “preprocessing,” “IBSI,” “CT,” “MRI,” and “PET.” Primary empirical studies, systematic and narrative reviews, and consensus guidelines were eligible if they addressed reproducible feature generation or methodological standardization. The findings informed a proposed \textit{minimum set of data elements} to represent the methodological context of quantitative features within mCODE. This set aims to improve methodological transparency and support the investigation of variability and bias, but does not itself guarantee harmonization or eliminate bias. The reviewed references are listed in Supplementary Tables S-1 and S-2, among which we particularly highlight the comprehensive literature review by Espinasse et al. (2020) \cite{Espinasse2020-yw} for its thorough synthesis of acquisition and reconstruction parameter effects, and Teng, X. et al. (2024)\cite{Teng2024-bv}, for their guideline recommendation to enhance transparency and reliability of radiomics studies. These studies were complemented by the IBSI online documentation \cite{ibsi2020docs}, which further details the essential medical image acquisition elements required to capture methodological context.

\vspace{-5mm}

\subsection{Designing the mCODE-based ontology}

We developed ORCHID by integrating mCODE STU4 with our proposed multimodal extension. Our ontology engineering process followed METHONTOLOGY, a structured methodology used to build ontologies from scratch, comprising domain and scope specification, reuse of existing ontologies, conceptual modelling, implementation, and evaluation \cite{Fernandez-Lopez1997-sy,Uschold1996-os}. METHONTOLOGY has been successfully applied in healthcare ontology development previously, including models for COVID-19, pneumonia, and healthcare systems \cite{Mcheick2022-ef,Vinson2022-ls,ZESHAN2012340}. The METHONTOLOGY workflow applied to this project is summarised in Figure~\ref{fig:methontology} and is described below.

\vspace{-5.5mm}
\subsubsection{Specification of the Ontology Domain and Scope}

ORCHID was designed to represent cancer diagnosis, treatment, and outcomes, with a focus on general oncology, radiation oncology, and medical physics. Its intended purpose is to support the integration and retrieval of heterogeneous clinical, imaging, and RT research data. Intended users include researchers, clinicians, and healthcare research institutions.

The scope was defined using ontology specification guidance from METHONTOLOGY and established ontology-development practices \cite{Mcheick2022-ef, Fernandez-Lopez1997-ea, Noy2001Ontology101}. The requirements addressed the ontology's domain, purpose, intended users and applications, and they included concepts and competency questions (CQs). These requirements are summarized in Table~\ref{tab:ontology-requirements}.

\begin{longtable}{>{\raggedright\arraybackslash}p{0.30\textwidth} >{\raggedright\arraybackslash}p{0.65\textwidth}}
\caption{Ontology requirements specification.}
\label{tab:ontology-requirements} \\
\toprule
\textbf{Question} & \textbf{Answer} \\
\midrule
\endfirsthead

\multicolumn{2}{c}{\tablename\ \thetable{} -- \textit{continued from previous page}} \\
\toprule
\textbf{Question} & \textbf{Answer} \\
\midrule
\endhead

\midrule
\multicolumn{2}{r}{\textit{Continued on next page}} \\
\endfoot

\bottomrule
\endlastfoot

(i) \textit{What is the domain that the ontology will cover?} &
ORCHID covers cancer diagnosis and treatment, with a particular focus on Oncology, Radiation Oncology, and Medical Physics. \\
\addlinespace

(ii) \textit{What is the purpose of the ontology?} &
ORCHID facilitates the integration, aggregation, and interoperability of heterogeneous clinical and research data. It represents patient information from diagnosis through treatment and outcomes, supporting predictive analyses and applications such as digital twins.\\
\addlinespace

(iii) \textit{Who are the intended users of the ontology?} &
The intended users of ORCHID are researchers, clinicians, and healthcare research institutions involved in Oncology, Radiation Oncology, and Medical Physics. \\
\addlinespace

(iv) \textit{What are the intended uses or applications of the ontology?} &
ORCHID is designed to support AI and data science workflows, RT outcome prediction, comparative population studies, and digital twin development, aligned with the mCODE objective of standardized data capture.\\
\addlinespace

(v) \textit{What concepts should be included from the ontology?} &
ORCHID's ontology concepts align with the mCODE terminology and specification \cite{mCODE_terminology}, including diagnosis and procedure codes from the International Classification of Diseases, 10th version (ICD-10)\cite{ncbo2026icd10cm} and International Classification of Diseases for Oncology, 3rd Edition (ICD-O-3); observations and laboratory tests from the Logical Observation Identifier Names and Codes (LOINC)\cite{regenstrief2026loinc}; cancer-related terminology from the National Cancer Institute Thesaurus (NCIt)\cite{fragoso2004overview}; medications from RxNorm\cite{1516084}; and genomic information represented using the Sequence Ontology (SO)\cite{Eilbeck2005-fl} and HUGO Gene Nomenclature Committee (HGNC\cite{Bruford2020-tk}). Clinical findings, disorders, body structures, qualifiers, and selected high-level imaging procedures are represented using Systematized Nomenclature of Medicine – Clinical Terms (SNOMED CT)\cite{vuokko2023systematized}. RT treatment concepts are further represented using the Operational Ontology for Oncology (O3)\cite{Mayo2023-tr}, while the extension incorporates RadLex\cite{Mejino2008-ip} and O-Raw\cite{Shi2019-xg}, which is aligned with IBSI.
\\
\addlinespace

(vi) \textit{What types of questions should the ontology be able to answer?} &
ORCHID should support identifying patients eligible for specific therapies or clinical trials, comparing outcomes across cohorts, and analyzing associations between patient attributes—including RT parameters, genomic variants, and biomarkers—and treatment responses and outcomes. Competency Questions (CQs) were therefore formulated to ensure that the ontology can answer both clinical decision-making and research-oriented queries in a semantically interoperable manner. These questions also guided the development of the conceptual models presented in the following sections.\\
\end{longtable}
\vspace{-5mm}
\subsubsection{Reuse of Existing Ontologies}
\label{reuse}
ORCHID reuses concepts from established biomedical ontologies and terminology standards to promote consistency with existing resources and support semantic interoperability. The selection of these resources was guided by the terminology bindings and \texttt{ValueSet} definitions specified in the mCODE STU4 Implementation Guide~\cite{HL7mCODESTU4_2025,HL7mCODEValueSets}, which associate individual data elements with permitted codes from standard terminology systems. For example, the \texttt{Primary Cancer Condition} data element is associated with a ValueSet containing codes from SNOMED CT and ICD-10-CM \cite{mcode_primary_malignant_neoplasm}.  Thus, as summarized in Table \ref{tab:ontology-requirements}, SNOMED CT was used to represent clinical concepts, including anatomical structures, clinical situations, findings, and observable entities, while maintaining mappings to the ICD-10 classification system. Laboratory terms and clinical observations were also drawn from SNOMED CT, whereas medication concepts were aligned with RxNorm, which provides standardized identifiers and semantic representations for clinical drugs, including ingredients, strengths, and dose forms. For genomics, concepts from HGNC, SO, LOINC, and NCIt controlled terminology \cite{GOLBECK200375} were incorporated.

\begin{figure}[h]
\vspace{-4mm}
\centering
{\captionv{12}{Short title - can be blank}{\small Ontology term reuse workflow using METHONTOLOGY. Existing biomedical ontology files (OWL/TTL) are processed via the ROBOT OBO tool to export target term subsets, which are subsequently imported into the main ORCHID ontology. OWL: Ontology Web Language, and TTL: Terse RDF Triple Language.\label{fig:existing_ontologies}}}
\includegraphics[width=0.80\linewidth]{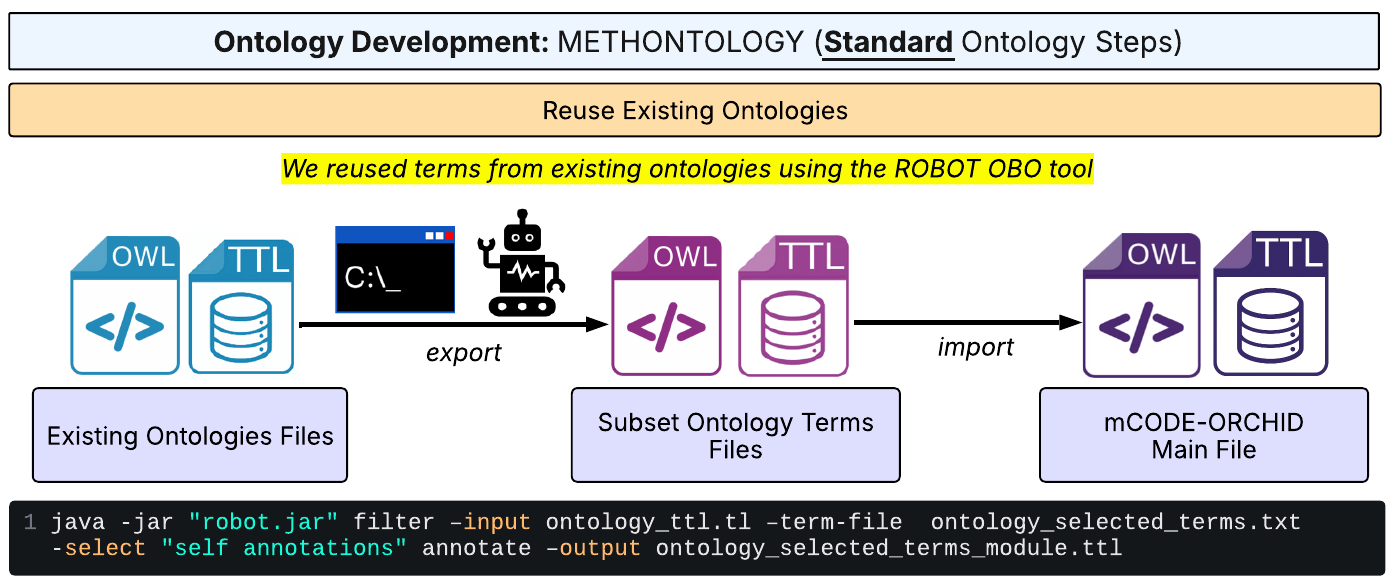}
\end{figure}

Imaging-related concepts were aligned with the RadLex ontology \cite{article_Radlex}, which provides a standardized vocabulary for radiology procedures, imaging modalities, and radiological findings. RadLex also facilitate the interpretation and harmonization of imaging metadata associated with DICOM header attributes, such as protocol names and imaging parameters \cite{RadLexPlaybookUserGuide,RadLexPlaybookSeries}. Radiomics concepts were represented using O-RAW Radiomics Ontology\cite{Shi2019-xg}, which models radiomics workflows, feature extraction processes, and quantitative imaging features derived from medical images. The Units of Measurement Ontology (UO) \cite{Gkoutos2012UO} was used to represent measurement units corresponding to Unified Code for Units of Measure (UCUM) codes, supporting consistency with the mCODE terminology.

To incorporate selected concepts from existing ontologies while limiting the size of the resulting ontology, we used ROBOT v1.9.10\cite{Jackson2019ROBOT}, a command-line tool and library for automating ontology development tasks, to extract relevant terms from source ontology files in Ontology Web Language (OWL) or Terse RDF Triple Language (TTL) format. Specifically, the \texttt{filter} command was used with the \texttt{--select} option and the \texttt{STAR} selection method to extract the target classes and their associated axioms from the source ontologies selected for reuse. When required, the ELK open-source reasoner was run beforehand to infer additional subclass relationships and support the extraction of the complete subclass hierarchy. This methodology has been described in previous ontology-development studies \cite{social_dOntologies, Ortega2023-uu, Matentzoglu2022-kw}. The resulting subsets extracted from the source ontologies were saved as a TTL or OWL files and subsequently imported into the main ORCHID ontology, as illustrated in Figure~\ref{fig:existing_ontologies}. 
\vspace{-5mm}

\subsubsection{Conceptual Modeling}

Conceptual models provide abstract representations of a domain's main concepts and the relationships between them. In ontology development, they serve as an intermediate step between domain analysis and formal implementation, organizing domain knowledge before it is encoded in a formal language \cite{Noy2001Ontology101, Guizzardi}. Accordingly, we developed conceptual models to represent the relationships among the ontology elements and guide their formalization. These models covered the three groups in extension, \textit{Medical Image, Radiomics}, and \textit{Dosiomics}, as well as the six mCODE thematic groups.

We developed competency questions (CQs) to guide ontology conceptualization and evaluate information retrieval. The initial CQs, adapted from El Ghosh et al. \cite{El_Ghosh2026-je}, addressed established clinical mCODE domains, including \textit{Disease Characterization}, \textit{Health Assessment, Cancer Treatments}, and \textit{Outcomes}. To evaluate the distinctive contributions of ORCHID, we added extension-specific CQs addressing relationships among clinical diagnoses, medical imaging studies, radiomics reports, genomic findings, RT dose calculations, and dosiomics reports. Example sets of CQs are provided in Supplementary Table S-3.

Cardinality constraints were included in the conceptual models to specify the number of instances permitted in each relationship, guiding ontology design and supporting consistency with mCODE. However, these multiplicities were not enforced as formal cardinality restrictions in OWL, allowing the relationships to remain flexible during ontology implementation. The multiplicities indicate exactly one (1), zero or one (0..1), one or more (1..*), or zero or more (0..*) instances\cite{halpin2005information}.
\vspace{-5mm}

\subsubsection{Ontology Implementation}

To make our ORCHID ontology publicly available to the research community in a \textit{standardized}, machine-readable format,  we implemented the ontology using Protégé v5.6.9, an open-source ontology editor developed at Stanford University \cite{Protege}. Using this software, we were able to create each class and relationship (object and data properties) encompassed by our designed conceptual models, as shown in Figure \ref{fig:methontology}. We then imported the subset ontology modules selected during the ontology reuse phase (Section \ref{reuse}).

Next, we enriched our ontology by creating metadata using RDF Schema (RDFS)\cite{brickley2014rdfs} and Simple Knowledge Orientation System (SKOS)\cite{Miles2009-bd}, i.e. \texttt{rdfs:label}, \texttt{skos:altlabel}, \texttt{skos:definition}, and \texttt{rdfs:comment}. We also added the data annotation \texttt{skos:exactMatch} to map equivalent concepts from other imported ontologies. This approach allowed us to designate a representation within ORCHID while retaining explicit mappings to corresponding concepts in external standards, without necessarily asserting formal OWL equivalence, which might introduce errors when running a reasoner \cite{Miles2009-bd, Matentzoglu2022-od}. To illustrate, we used \texttt{skos:exactMatch} to document mappings between the \texttt{Imaging Modality} concepts represented in RadLex and O-RAW, as well as between concepts in SNOMED CT and LOINC, such as \texttt{Body Height}. However, when concepts were determined to be semantically equivalent, we retained formal OWL equivalence (\texttt{owl:equivalentClass}), and included these equivalence axioms directly within the ontology. For example, the \texttt{Computed Tomography} classes from RadLex and the National Cancer Institute Thesaurus (NCIt) were declared equivalent using \texttt{owl}. Together, these approaches facilitate the integration of heterogeneous data sources when creating instances, by supporting cross-terminology mappings while reserving formal equivalence axioms for concepts whose meanings are sufficiently aligned. Finally, we consolidated the ontology and its imported modules into a single file; we used the \texttt{Merge ontologies} option under Protégé's \texttt{Refactor} menu. We saved the resulting ontology in OWL, JSON, and TTL formats. 
\vspace{-5mm}

\subsubsection{Technical Evaluation}
Once the ORCHID was developed, the technical evaluation followed a two-phase validation approach: (1) structural validation and (2) instance population and functional evaluation (see Figure \ref{fig:methontology}). In the first phase, we assessed the ontology's design, logical structure, and semantics by reviewing its class hierarchy, properties, and relationships against the conceptual models. We then evaluated its logical consistency using the HermiT reasoner (v1.4.3) to identify logical contradictions and unsatisfiable classes and to examine the consistency of property domains and ranges\cite{Chatterjee2021-se}. Additionally, we used ROBOT  for ontology validation and quality assurance, using the command line \texttt{report}.

In the second phase, after validating the ontology schema, we manually populated it with ten random instances using clinical information from the mCODE-based Synthea resource \cite{hl7_mcode_test_data}, genomics data from examples available on the mCODE website \cite {genomics-example}, and mock data values for our extension. Next, we performed functional evaluation by verifying CQs formulated to assess whether ORCHID could represent and retrieve the information required by its intended use cases. We evaluated three extension-specific questions addressing: (1) links between cancer diagnoses and clinical information; (2) links between RT courses, dose-calculation information, and dosiomics reports; and (3) retrieval of genomics, radiomics, and dosiomics information within a shared patient context. We translated these questions into SPARQL queries and executed them against the populated ontology using Protégé’s Snap SPARQL Query Tab to verify whether the expected information could be retrieved\cite{Shi2019-xg}. 
\vspace{-5mm}

\subsection{Web Browser Development}

To facilitate access to and exploration of the ORCHID ontology, we designed an interactive, web-based browser intended to enable users to navigate its components, including classes, exact matches, equivalencies, object properties, and data properties. We aimed to follow the same design structure as the \href{https://snomedbrowser.org/?perspective=full&conceptId1=404684003&edition=MAIN/2026-10-01&release=&languages=en}{SNOMED CT Browser} and \href{https://radlex.org/}{RadLex Term Browser} web pages, and provide a hierarchical view of the ontology, facilitating the exploration of its structure and semantic relationships, as well as a search function for retrieving specific ontology elements. Additionally, we envisioned incorporating an interactive, tree-based visualization inspired by the reusable web visualization developed for the EDAM ontology \cite{edam} (adapted from \href{https://github.com/edamontology/edam-browser}{edam-browser} distributed under the MIT License), along with a feedback mechanism to allow users to submit suggestions for future improvements.
\vspace{-5mm}

\section{Results}
\vspace{-2mm}
\subsection{mCODE Extension}

\begin{figure}[h]
\centering
\includegraphics[width=1\linewidth]{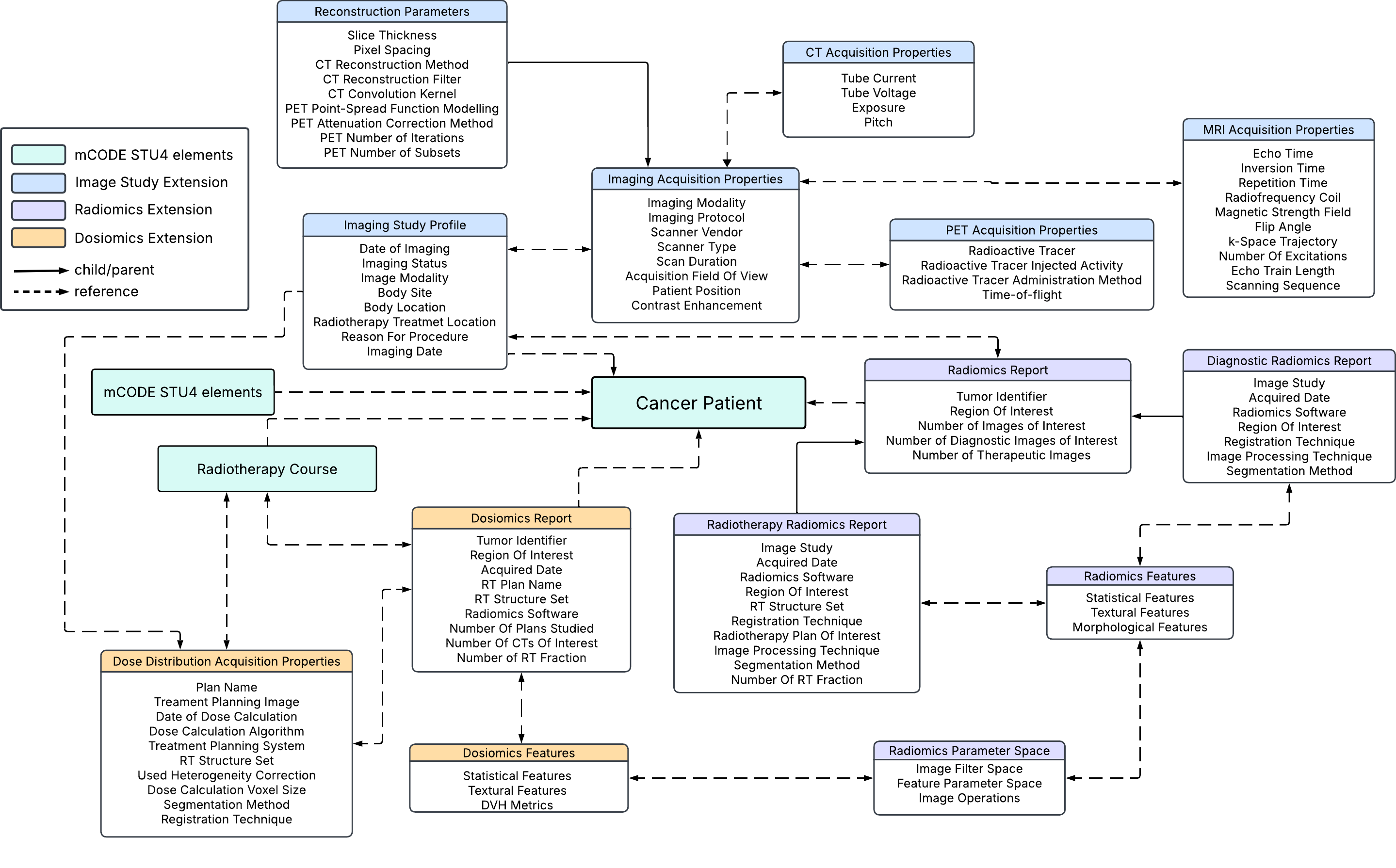}
\vspace{-16mm}
{\captionv{10}{Short title - can be blank}{\small Diagram of our mCODE extension for image study, radiomics, and dosiomics, using the mCODE structure formalism from \href{https://build.fhir.org/ig/HL7/fhir-mCODE-ig/}{mCODE STU4 version}. DVH: Dose-Volume Histogram; RT: Radiotherapy; CT: Computed Tomography; MRI: Magnetic Resonance Imaging; PET: Positron Emission Tomography; STU: Standard for Trial Use. \label{fig:mcodeextension_results}}}
\end{figure}

Based on the rapid literature review and following IBSI guidelines, we developed our proposed mCODE extension, shown in Figure \ref{fig:mcodeextension_results}. The extension resulted in 14 sets of data elements, encompassing a total of 99 elements. Within this framework, the medical imaging and radiomics components, which cover both diagnostic and RT images, comprise 11 sets containing 77 elements, while the dosiomics extension comprises 3 sets with 22 elements. Together, these extensions expand the scope of mCODE to support the integration of EHR data with imaging- and dose-derived biomarkers.
\vspace{2mm}

Within our extension, the \textit{Imaging Study Profile} represents imaging studies and their associated \textit{Imaging Acquisition Parameters}, which describe imaging protocols and scanner configurations, including slice thickness, pixel spacing, and reconstruction algorithms. The \textit{Radiomics Group} links radiomics reports to tumor-specific identifiers, accommodating patients with multiple primary tumors or disease sites and multiple associated imaging studies and analyses; it also supports specific-target analyses via a Region of Interest (such as treatment target volumes, organs, and organs-at-risk), from which the radiomics features are calculated. Thus, the \textit{Radiomics Report} references the Number of Images Of Interest (diagnostic and therapeutic images) associated with the analysis target. A similar structure was designed for a patient’s \textit{Dosiomics Report} and each \textit{Radiotherapy Course}. We extended the \textit{Radiotherapy Course} information to include and link relevant information to the \textit{Dosiomics Report}, such as how the dose distribution was calculated. We included elements to support adaptive RT (ART) workflows, enabling explicit tracking of RT plans and replans.

\vspace{-5mm}
\subsection{ORCHID Development}
\subsubsection{Ontology Design and Implementation}
Based on the conceptual models for the proposed mCODE extension presented in Figure \ref{fig:map_extension}, and existing mCODE group elements (Figures S-1—S-5 in the Supplementary Material), we defined the ontology's classes, object properties, and datatype properties to represent domain concepts, their relationships, and their attributes. In the conceptual maps, in addition to each mCODE and extension element, we also show auxiliary classes that we created to link to higher hierarchy levels.

\begin{figure}[h!]
    \centering
    {\captionv{12}{Short title - can be blank}{\small High-level conceptual map illustrating key classes of the mCODE's groups and our extension. To improve readability and accommodate space constraints, this diagram presents a partial representation of the core architecture.  \label{fig:map_extension}}}
\includegraphics[width=1\linewidth]{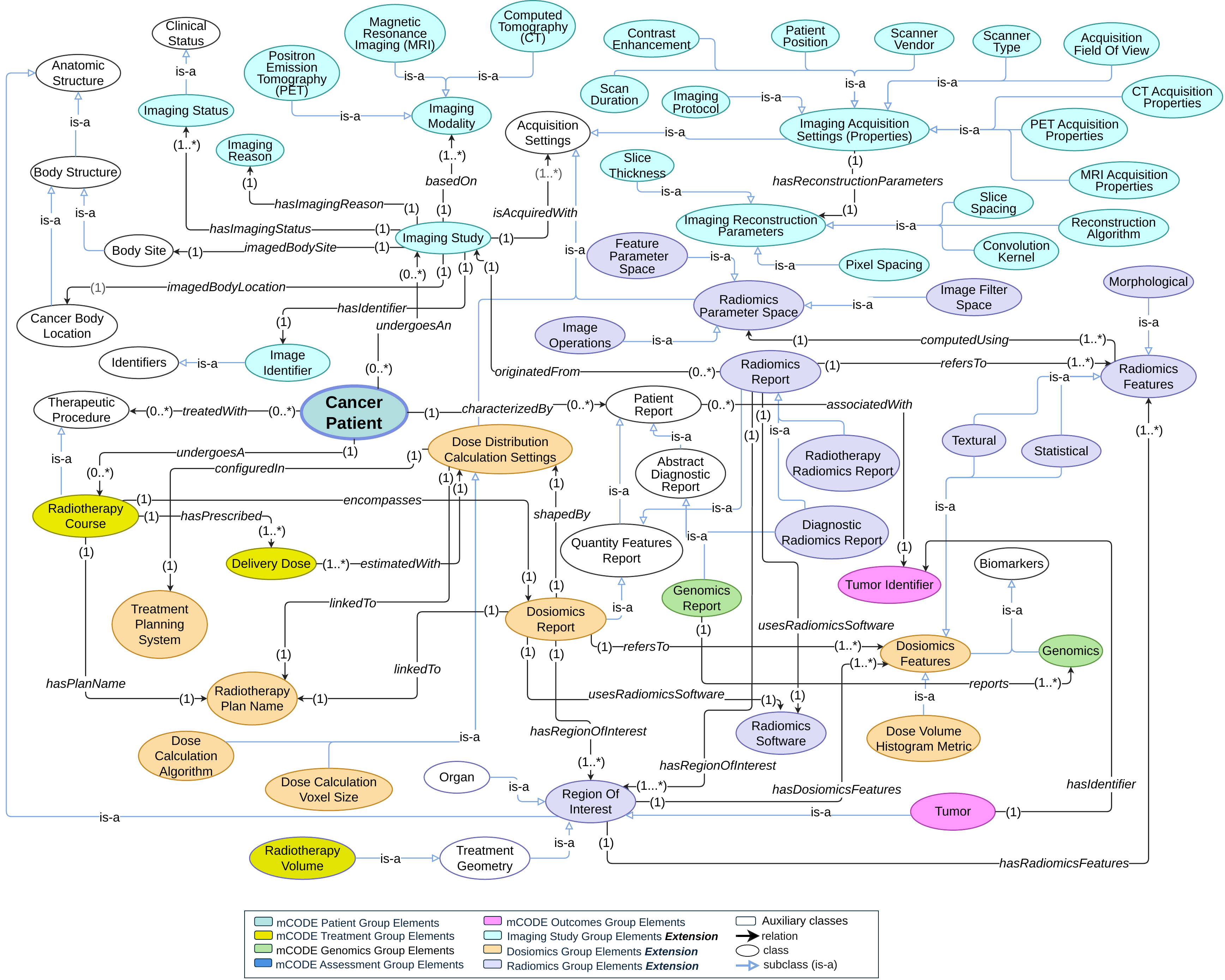}
\end{figure}

We mapped our conceptual models, our described classes and relationships, into Protégé v5.6.9 on an Intel Core i7 processor and 32 GB RAM, with 16 GB allocated to the Java Virtual Machine (JVM). Figure S-6 in the Supplementary Material summarizes our ontology implementation workflow. Table S-4 in the Supplementary Material summarizes the metrics results from our ontology. 

\begin{itemize}[leftmargin=0.5em, labelsep=0.3em]
\item \textbf{Classes:} ORCHID comprises 254 classes, excluding imported classes, and 152,129 classes when imported terms are included. These classes represent concepts from the mCODE framework and the proposed extension, including \texttt{Annotation}, \texttt{Anatomic Descriptor}, \texttt{Anatomic Structure}, \texttt{Biomarkers}, \texttt{Health Condition}, \texttt{Methods or Techniques}, \texttt{Patient Report}, \texttt{Processing And Acquisition Settings}, and \texttt{Product}. These classes support the representation of clinical information, anatomical structures, genomics-, imaging-, and dose-derived biomarkers, and associated acquisition and processing settings. Key subclasses include \texttt{Genomics Report, Radiomics Report}, and \texttt{Dosiomics Report} under \texttt{Patient Report}; and \texttt{Radiotherapy Treatment Location}, \texttt{Treatment Geometry}, and \texttt{Gene} under \texttt{Anatomic Structure}.\\

\item \textbf{Object and Datatype properties:} The current version of the ontology (release version 1.0.0, November 2026) includes 224 newly defined object properties and 352 in total, including imported object properties. Representative object properties include \texttt{diagnosedWith} (Domain: \texttt{Cancer Patient}, Range: \texttt{Primary Malignant Neoplastic Disease}); \texttt{deliveredTo} (Domain: \texttt{Radiotherapy Delivery Dose}, Range: \texttt{Radiotherapy Volume}); \texttt{imagedBodyLocation} (Domain: \texttt{Imaging Study}, Range: \texttt{Cancer Body Location}), and \texttt{configuredIn} (Domain: \texttt{Dose Distribution Calculation Setting}, Range: \texttt{Treatment Planning System}). The ontology currently includes 47 datatype properties, excluding imported datatypes; additional properties can be added in future releases depending on scientific community feedback. Representative properties include \texttt{hasStringValue}, which associates an instance of \texttt{Identifiers} with a string value (\texttt{xsd:string}); \texttt{hasImagingDate}, which records the date and time of an \texttt{Imaging Study} (\texttt{xsd:dateTime}); and \texttt{hasStatusDate}, which records the date and time associated with a \texttt{Clinical Status} (\texttt{xsd:dateTime}).\\

\item \textbf{Imaging and biomarkers representation:} Each \texttt{Cancer Patient} is linked to one or more \texttt{Imaging Studies} through \texttt{undergoesAn}, characterized by their \texttt{Imaging Status}, \texttt{Imaging Reason}, and \texttt{Imaging Modality} (MRI, CT, and PET). Imaging studies are connected to acquisition and reconstruction settings, capturing scanner, acquisition, and reconstruction parameters. Imaging terminology follows RadLex, NCIt, and O-RAW. Radiomics information is represented through \texttt{Radiomics Reports}, including \texttt{Diagnostic Radiomics Reports} and \texttt{Radiotherapy Radiomics Reports}, which are linked to source images through \texttt{originatedFrom}, tumor identifiers through \texttt{hasIdentifier}, and anatomical or treatment-related regions through \texttt{hasRegionOfInterest}. Extracted features are linked with reports through the \texttt{RegionOfInterest} class and through the \texttt{hasRadiomicsFeatures} and \texttt{reportsTo} object properties, and computed using a \texttt{Radiomics Parameter Space} and specified \texttt{Radiomics Software}. Similarly, \texttt{Dosiomics Reports} are linked to \texttt{Radiotherapy Courses}, \texttt{Dose Distribution Calculation Settings}, and \texttt{Regions Of Interest}, with resulting features connected through \texttt{hasDosiomicsFeatures}. Dose calculation parameters include the algorithm and voxel size, while dose-derived features include dose-volume histogram metrics. Genomic information is captured through the \texttt{Genomics Report}, based on analyses of human specimens, typically obtained from tumor biopsies or other biological samples \cite{Costain2021-hu,Roychowdhury2014-zk}. Findings are represented as \texttt{Genomic Observations}, including variant components linked to the affected \texttt{Gene} and genomic region components specifying variant coordinates and reference coordinate systems. The model supports HGVS-compliant descriptions of coding changes, molecular consequences, allelic state and frequency, copy number type, and cytogenetic location.
\end{itemize}
\vspace{-5mm}

\subsubsection{ORCHID Evaluation}

Once the ontology was implemented in the ontology editor software, logical consistency was verified using the HermiT reasoner v1.4.3. Classification took 1303 seconds (about 22 minutes). The first run reported no inconsistency and no unsatisfiable classes (no classes subsumed by owl: Nothing). Quality assurance evaluation using ROBOT v1.9.10 identified three missing ontology metadata annotations (description, license, and title), which were subsequently added. Restricting the ROBOT report to the custom IRI namespace yielded 488 missing-definition flags, considered false positives because definitions were initially represented using SKOS annotations rather than the OBO-standard \texttt{IAO\_0000115} property. We subsequently added definitions for the ontology's classes; object and datatype properties remained undefined since their labels were self-explanatory.

For the second evaluation, as a proof-of-concept, we manually created ten breast cancer patient instances and developed three SPARQL query scripts to answer three data-retrieval CQs: \textit{``Which blood pressure values had patients with cancer stage below III?''}, \textit{``Which treatment planning system and dose calculation algorithm were used to compute the dose associated with each dosiomics report?''}, and \textit{``What were the kurtosis value and the Dmax for the PTV of breast cancer patients with a genomic variant located in the BRCA1 gene?''}. Results for the first two queries are provided in Figures S-7 and S-8 of the supplementary material, while Figure \ref{fig:evaluation} reports the results for the final query. Thus, the reasoning process confirmed that the ontology is logically consistent, allowing for automated classification and error-free execution of research-focused queries.

\begin{figure}[h!]
    \centering
    {\captionv{12}{Short title - can be blank}{\small SPARQL results from our research question query: \textit{``What were the kurtosis value and the Dmax for the PTV of patients with a genomic variant located in the BRCA1 gene?''}. On the left is the SPARQL query that retrieves the patients with a genomic variant present in any gene. On the top-right is the output table for this query; in the middle-right are the results after filtering for the ``BRCA1'' gene, retrieving the correct result; and at the bottom-right are the remaining patients out of the ten instances created.\label{fig:evaluation}}}
\includegraphics[width=1\linewidth]{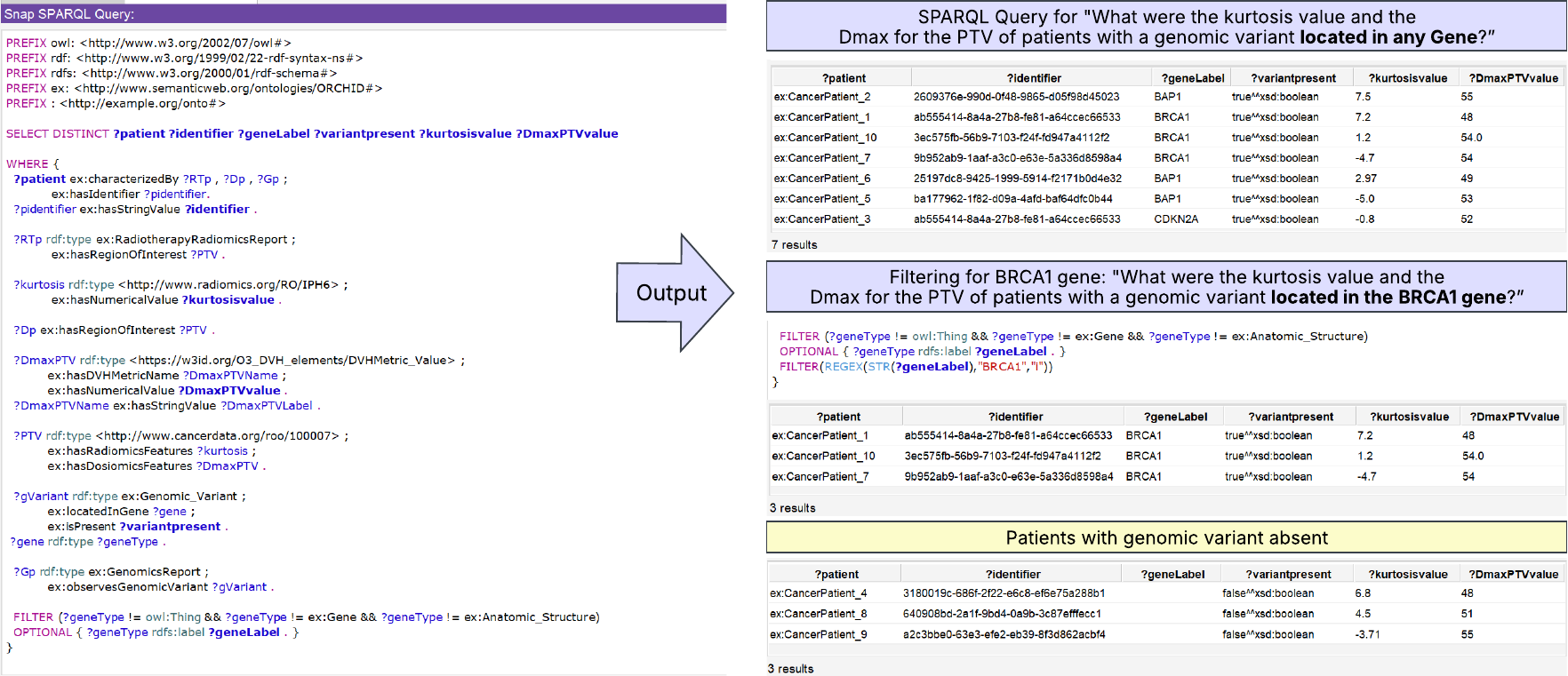}
\end{figure}
\vspace{-5mm}
\subsection{ORCHID Web Browser}
We successfully developed the \href{orchid-browser.kildealab.com}{ORCHID Taxonomy Browser} to facilitate the interpretation and exploration of the ontology, which is currently publicly available. Using this browser, each element can be inspected to access its associated information in a tree format, including its label, alternative labels, definition, superclasses, subclasses, and relationships (object and datatype properties). We included free text boxes where users can provide feedback. We also implemented the associated EDAM ontology interactive view and added a Relationship Map, where we interactively display a short range of the associated conceptual map. Using this website, users can access ORCHID-associated documentation and follow up on updated versions.

\vspace{-5mm}

\section{Discussion}

Multicenter datasets are essential for developing robust, generalizable AI models in oncology, yet differences in EHR systems, data structures, terminology, and imaging and RT workflows hinder cross-institutional data integration and harmonization \cite{Kelly2019-os,Matta2024-dl,Nada2026-vj}. These challenges are particularly relevant to radiomics and dosiomics, whose reproducibility depends on patient and treatment characteristics, image acquisition and reconstruction, preprocessing, feature extraction, and dose calculation information, which are often difficult to extract and harmonize \cite{articleAI2,Traverso2018-rg}. To address this gap, we first extended mCODE STU4 by introducing dedicated data elements for medical imaging, radiomics, and dosiomics, and subsequently developed ORCHID: Ontology for Radiomics, Clinical Health, and Imaging Data, an OWL-based semantic framework that retains mCODE's clinical foundation and incorporates our extension. The extension links imaging acquisition and reconstruction parameters, radiomics feature-extraction settings, and quantitative features derived from RT dose distributions to relevant patient, imaging, and treatment information, providing the context needed to interpret these features.

To support semantic interoperability, ORCHID reuses concepts from established ontologies, including O3, RadLex, and O-RAW, building on existing representations rather than redefining established oncology concepts. While O3 covers a broad range of oncology and RT concepts, ORCHID focuses on a \textit{minimum} set of data elements needed to represent the methodological context of quantitative imaging and dose-derived features. This focused extension may help researchers identify methodological differences, investigate potential sources of variability, and support the development of future harmonization workflows for radiomics and dosiomics research. A comprehensive comparison of existing oncology and radiomics data elements representation frameworks with ORCHID is presented in Table S-5 in the Supplementary Material.  

Recent work by El Ghosh et al. (2026) \cite{El_Ghosh2026-je} introduced the Cancer Image Europe Hyperontology, a broad framework for integrating heterogeneous clinical, biological, and imaging data across healthcare and imaging standards. Like ORCHID, their approach draws on established ontological resources and incorporates mCODE elements to support semantic interoperability. However, the two approaches differ in scope and focus. The Hyperontology supports broad integration across oncology and medical imaging, whereas ORCHID extends mCODE STU4 with a minimum set of data elements representing medical imaging acquisition and reconstruction, radiomics feature extraction, and dosiomics in the context of cancer diagnosis and RT, including dose-calculation parameters. Although the Hyperontology's competency questions cover selected aspects of diagnosis, imaging, and treatment, they do not explicitly demonstrate retrieval of radiomics feature-extraction parameters or dosiomics features linked to RT dose calculations. In contrast, ORCHID includes extension-specific competency questions to assess retrieval across clinical, imaging, radiomics, and dosiomics domains. By linking quantitative features to their clinical and methodological contexts, ORCHID may help researchers investigate sources of variability and bias. It is intended to complement broader oncology ontologies and may be integrated into existing semantic frameworks for radiomics and dosiomics research. 

Additionally, we published a web-based browser to facilitate exploration of ORCHID, its elements and relationships, making its structure more accessible to researchers. Explicitly representing these parameters may support reproducibility by making methodological differences easier to identify. Our ontology serves as a foundation for knowledge bases that connect clinical records, imaging studies, radiomics reports, dosiomics reports, and RT information. Such knowledge bases may facilitate patient-level data retrieval and the preparation of datasets for multicenter research and AI development. However, an ontology alone does not guarantee data harmonization or equivalence between features generated by different pipelines.

This study provides an initial proof-of-concept, but evaluation was limited to ten synthetic patient instances, and the narrative literature review may have missed relevant studies or data elements. Real-world interoperability and harmonization have not yet been established. Although the ontology cannot itself eliminate bias or ensure equivalence between features generated by different pipelines, explicit representation of feature-generation parameters may help researchers identify methodological variability and investigate its contribution to bias in AI models.

Future work will evaluate ORCHID using larger, real-world, multicenter datasets, develop a data-mapping and preprocessing module to populate the ontology, alongside feature-extraction and harmonization modules. We plan to develop a cancer-specific knowledge base and investigate ontology-based reasoning. These efforts will assess the extension's practical contribution to interoperable biomarker data, reproducible radiomics and dosiomics research, and the development of more robust AI models in oncology.
\vspace{-5mm}

\section{Conclusions}
We developed ORCHID: Ontology for Radiomics, Clinical Health, and Imaging Data, an mCODE-based ontology that integrates clinical oncology data with medical imaging, radiomics, and dosiomics. To our knowledge, this is the first attempt to extend mCODE with dedicated representations for both radiomics and dosiomics. By linking these data to clinical, diagnostic, and RT contexts, ORCHID provides a foundation for semantic data integration and patient-level information retrieval, with potential applications in multicenter research and AI-driven radiation oncology. Further validation using real-world datasets is needed.
\vspace{-5mm}

\section*{Authors Contribution}

Odette Rios-Ibacache: Conceptualization; methodology and software; formal analysis; manuscript writing; manuscript editing. Gaetan Butault: Methodology and manuscript editing. Amal Zouaq: Conceptualization, supervision, manuscript editing, and funding acquisition. John Kildea: Conceptualization; formal analysis; manuscript writing; manuscript editing; supervision; funding acquisition. All authors have read and agreed to the published version of the manuscript.
\vspace{-5mm}
\section*{Conflict of Interest Statement}
The authors have no conflicts of interest to declare.
\vspace{-5mm}
\section*{Acknowledgments}

The authors acknowledge that mCODE\textregistered{} is a registered trademark of the American Society of Clinical Oncology (ASCO). Reference to mCODE\textregistered{} in this work is for academic identification and interoperability purposes only and does not imply endorsement, sponsorship, or affiliation by ASCO. The authors acknowledge Ningyuan Li for lending his 32 GB laptop for this research, Luc Galarneau for his technical support with website development, and the Digital Research Alliance of Canada for hosting the website.

This research was supported by the Fonds de recherche du Québec–Santé (FRQS) Dual Chair in AI and Digital Health held by Prof. John Kildea, and doctoral fellowships from AAPM and FRQS held by Odette Rios-Ibacache. 
\vspace{-5mm}
\section*{Copyright and Licensing}
\copyright~2026 Odette Rios-Ibacache, Gaetan Butault, Amal Zouaq, and John Kildea. ORCHID: Ontology for Radiomics, Health, and Imaging Data work is licensed under a Creative Commons Attribution-ShareAlike 4.0 International License (\textbf{CC BY-SA 4.0}). To view a copy of this license, visit \url{https://creativecommons.org/licenses/by-sa/4.0/}.
\vspace{-5mm}
\section*{Use of Generative AI Tools}
 Google Gemini (free web version, accessed September–October 2026) was used for manuscript editing assistance in grammar and flow only. Claude (Anthropic, Sonnet 5.5, September 2026) was used to assist in drafting and debugging JavaScript code for our interactive website. The authors reviewed, edited, and assume full responsibility for all code and manuscript text. The ontology, research design, and scientific content were developed entirely by the authors. 
 \vspace{-5mm}
\section*{References}
\footnotesize
\addcontentsline{toc}{section}{\numberline{} References}
\setstretch{0.5}
\vspace*{-5mm}
\bibliography{./example} 
\bibliographystyle{./medphy.bst} 




\newpage
\includepdf[pages=-]{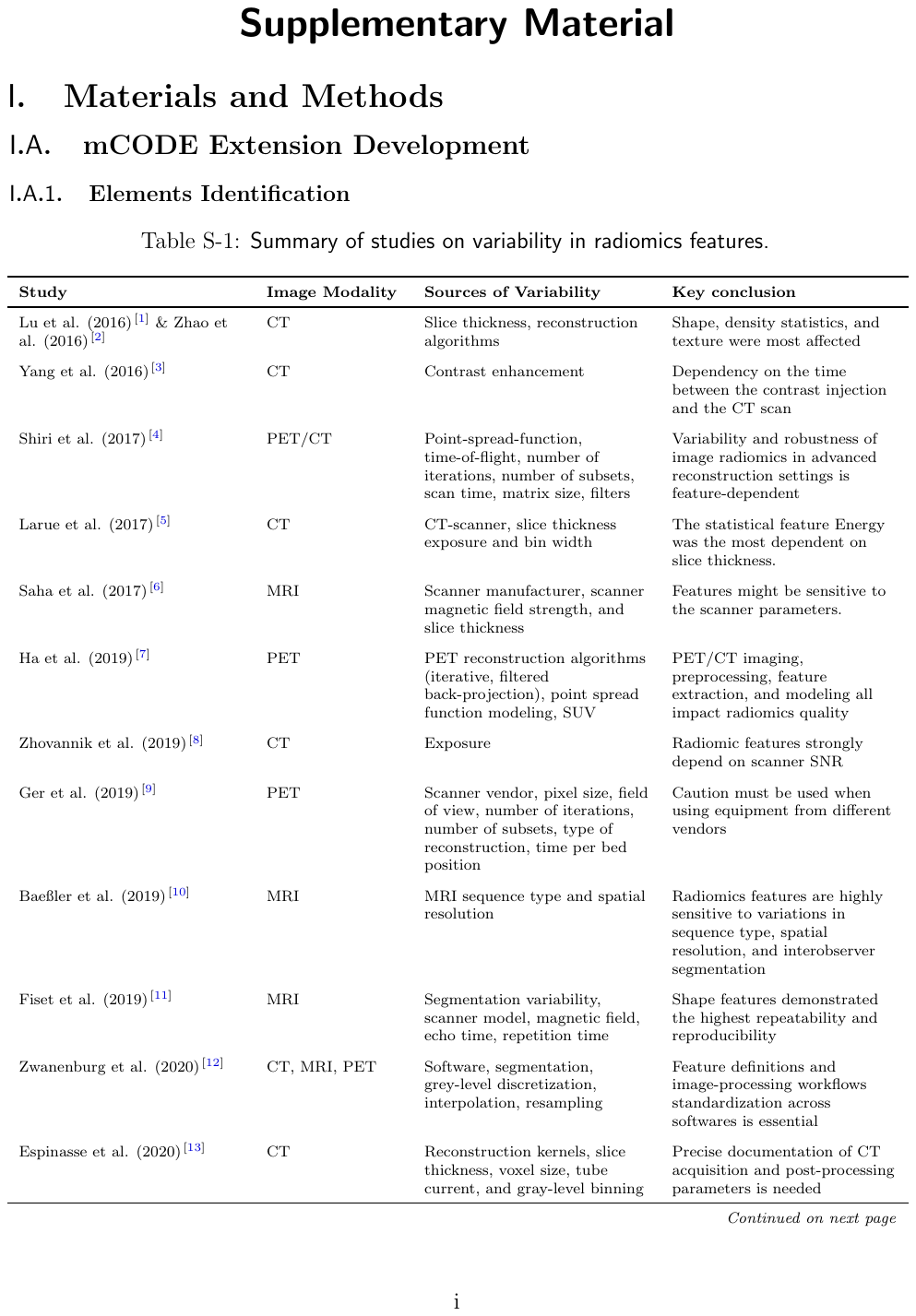}

\end{document}